\documentclass{aa}
\usepackage{txfonts}
\begin{document}

\def\teff{${\rm T_{\rm eff}}$}
\def\gr{$\log g$}
\def\alin{${\rm A(\rm Li})_{\rm NLTE}$}
\def\ali{${\rm A(\rm Li)}$}
\def\ciso{${\rm ^{12}C/^{13}C}$}

\title{
Discovery of a thin lithium plateau \\ among metal-poor red giant branch stars
\thanks{Based on observations collected at the ESO-VLT under programs 
68.D-0546, 69.D-0065, 70.D-0009, 71.B-0529, 072.B-0585, 074.B-0639, 076.D-0451, 078.B-0238,  
090.B-0605, 092.D-0742, 099.D-0287, 0103.D-0310, 0104.B-0487, 0104.D-0059,
165.N-0276, 169.D-0473, 170.D-0010, 281.D-5015, 380.D-0040, at the La Silla 
Observatory under the program 60.A-9700, at the Magellan telescope under programs CN2017A-33 
and CN2017B-54, and on data available in the ELODIE archive.}
}

\author{A. Mucciarelli\inst{1,2}, L. Monaco\inst{3}, P. Bonifacio\inst{4}, M. Salaris\inst{5,6}, M. Deal\inst{7}, M. Spite\inst{4}, O. A. Richard\inst{8}, R. Lallement\inst{4}}
\offprints{A. Mucciarelli}

\authorrunning{A. Mucciarelli et al.}
\titlerunning{Li abundance in metal-poor RGB stars}

\institute{
Dipartimento di Fisica e Astronomia, Universit\`a degli Studi di Bologna, Via Gobetti 93/2, I-40129 Bologna, Italy
\and
INAF - Osservatorio di Astrofisica e Scienza dello Spazio di Bologna, Via Gobetti 93/3, I-40129 Bologna, Italy
\and
Departamento de Ciencias Fisicas, Universidad Andres Bello, Fernandez Concha 700, Las Condes, Santiago, Chile
\and
GEPI, Observatoire de Paris, Universit{\'e} PSL, CNRS, Place Jules
Janssen, 92195 Meudon, France
\and
Astrophysics Research Institute, Liverpool John Moores University, 146 Brownlow Hill, Liverpool L3 5RF, United Kingdom
\and
INAF - Osservatorio Astronomico d'Abruzzo, via M. Maggini, 64100, Teramo, Italy
\and
Instituto de Astrof{\'i}sica e Ci{\^e}ncias do Espa\c co, Universidade do Porto, 
CAUP, Rua das Estrelas, PT4150-762 Porto, Portugal
\and
Laboratoire Univers et Particules de Montpellier, Universit{\'e} de Montpellier, CNRS, Place Eug{\`e}ne Bataillon, 34095 Montpellier,
France
}

\date{Accepted by A\&A }

\abstract
{The surface lithium abundance \ali\ of warm metal-poor dwarf stars
exhibits a narrow plateau down to [Fe/H]$\sim$--2.8 dex, while 
at lower metallicities the average value drops by 0.3 dex with a significant star-by-star scatter
(called {\sl lithium meltdown}). This behaviour is in conflict with predictions of standard 
stellar evolution models calculated with the initial A(Li) provided by the standard Big Bang nucleosynthesis.

The lower red giant branch (LRGB) stars provide a complementary tool to understand 
the initial \ali\ distribution in metal-poor stars.
We have collected a sample of high-resolution spectra of 58 LRGB stars spanning a range of [Fe/H] between $\sim$--7.0 dex and $\sim$--1.3 dex. 
The LRGB stars display an \ali\ distribution clearly different from that of the dwarfs, 
without signatures of a {\sl meltdown} and with two distinct components: 
(a)~a thin \ali\ plateau with an average \ali\ =~1.09$\pm$0.01 dex ($\sigma$=~0.07 dex), and (b)~a small fraction of Li-poor stars  with \ali\ lower than $\sim$0.7 dex.

The \ali\ distribution observed in LRGB stars can be reconciled with 
an initial abundance close to the cosmological value, 
by including an additional chemical element transport in stellar evolution models. 
The required efficiency of this transport allows us to match also the Spite plateau 
lithium abundance measured in the dwarfs.

The emerging scenario is that all metal-poor stars formed with the same initial \ali\ 
but those that are likely the product of coalescence or that experienced binary mass transfer and show lower \ali . 
We conclude that \ali\ in LRGB stars is qualitatively compatible with the cosmological \ali\ value and that 
the {\sl meltdown} observed in dwarf stars does not reflect a real drop of the abundance at birth.
}


\maketitle
%
%________________________________________________________________

\section{Introduction}

After about four decades, the distribution of the surface lithium abundance, A(Li)\footnote{A(Li)=$\log{\frac{N_{Li}}{N_{H}}}+12$, 
where $N_{a}$ is the number fraction of element $a$.}, 
in Galactic halo dwarf stars still poses a challenge to stellar evolution models. 
Stars with effective temperature (\teff ) above $\sim$5600~K and [Fe/H] between $\sim -$2.8 dex and 
$\sim-$1.0 dex, display a constant level of surface lithium abundance  
(A(Li)$\sim$2.1-2.3 dex, the exact value depending on the adopted \teff\ scale)
with a very small dispersion, the so-called Spite plateau \citep[][]{spite82}.  
Initially interpreted as the primordial Li abundance synthesised during the Big Bang, this abundance level was 
later found to be three to four times lower than the predictions 
of standard big bang nucleosynthesis (SBBN) for the baryon-to-photon ratio determined by the 
WMAP and PLANCK satellites \citep[see][and references therein]{coc14}. 
This discrepancy is often referred to as the {\sl cosmological lithium problem}, and solutions were searched along
different routes, including nuclear physics, stellar physics as well as modifications  
to the SBBN \citep[see][for a review]{spite12}.

Observations have later shown that the Spite plateau crumbles 
below [Fe/H]$\sim$--2.8 dex \citep{asplund06,bonifacio07,aoki09,sbordone10,aguado19}. 
At these metallicities the mean A(Li) in warm dwarfs progressively decreases and the abundance 
dispersion becomes substantial, although
three stars with Spite plateau abundances have been observed \citep{bonifacio18,aguado19}.
This drop of the mean A(Li) is usually named {\sl lithium meltdown} and 
its origin is still unclear. It can be the signature of an inhomogeneous interstellar medium and varying levels  
of processing of primordial material by massive stars in the early universe, 
but it may also arise from depletion mechanisms inside the star or at the stellar surface,
acting differently in this metallicity regime than at higher [Fe/H]. 

The observed behaviour of A(Li) with [Fe/H] is problematic for standard stellar models that 
include convection and atomic diffusion (taking into account radiative accelerations)
as the only element transport mechanisms. 
The effect of atomic diffusion on the surface chemical composition 
is metallicity-dependent, becoming at a given age more efficient at lower metallicity 
where turn-off stars have an higher \teff\ because of the thinner (in mass) outer convective layers. 
This means that, assuming the same initial A(Li) for all metal-poor stars,
the surface A(Li) during the main sequence (MS) evolution should  
progressively decrease with decreasing initial [Fe/H] and also increasing \teff\ at constant initial [Fe/H],
at variance with the measured uniform A(Li).

The situation is further complicated by the fact that atomic diffusion is predicted to also deplete the surface
[Fe/H] during the MS evolution of these objects for initial [Fe/H] higher than $\sim -2.3$ dex,
and increase it for lower initial [Fe/H] due to the selective effect of radiative accelerations \citep{ric02}.
At any rate, a Li abundance plateau as a function of
the actual [Fe/H] is not predicted by stellar models including fully efficient atomic diffusion.

A natural inference from this discrepancy is that some additional element transport mechanism should occur 
during the MS evolution, to counterbalance the effect of
atomic diffusion and create a plateau for actual [Fe/H] values above
$\sim -$2.8 dex.  
This inference is confirmed also by the very similar abundances of various heavy elements measured 
in dwarf and RGB stars of some Galactic globular clusters \citep[see, e.g.,][]{korn06, mucciarelli11}.

At variance with atomic diffusion, which is a well established process derived from first principles and without free parameters 
\citep[see e.g. ][]{michaud15,salaris17}, there is yet no established physical description for these additional mechanisms. Mass loss \citep{vauclair95,vick13}  or rotation-induced mixing and penetrative convection \citep{dumont21}  can modify the surface A(Li),  and \citet{deal21b} showed that they are good candidates to explain the surface A(Li) of Population II stars. However, their modelling is still subject to sizeable uncertainties.
On the other hand, \citet{ric02} proposed as a pragmatic solution the inclusion of a turbulent diffusion coefficient\footnote{Its simple formulation could be easily implemented in any stellar evolutionary code.} able to partially counteract atomic diffusion in the outer layers of models for Spite plateau stars. 
Depending on its efficiency, this turbulent diffusion can even transport  
some extra lithium down to the Li-burning region ($T>2.5\times10^{6}$ K).

The added turbulent diffusion coefficient has been tuned to reproduce the measured abundances of 
few metals and A(Li) in MS 
and sub-giant branch stars in a handful of Galactic globular clusters \citep[see e.g.][]{korn06, gavel21} 
and the Spite plateau in field halo stars \citep{ric05, deal21b}, and 
reconcile the measured A(Li) with the SBBN Li abundance at a level of 1-2 $\sigma$ \citep{korn20}.
However, the lack of an established physical process for this turbulence   
reduces the predictive power of these models. 
Also, these models cannot explain the {\sl meltdown} observed at [Fe/H]$<$--2.8 dex.

A complementary view of the initial A(Li) in metal-poor stars is provided by 
chemical abundances in red giant branch (RGB) stars 
located after the completion of the first dredge-up (FDU) and below 
the RGB-bump luminosity level (these stars are denoted as lower RGB --- LRGB --- stars). 
Their surface \ali\  is predicted to depend on how much Li has been left in their
interiors after the MS, and
the maximum extension (in mass) reached by convection during the FDU. It turns out to be --for a given initial A(Li)-- 
very weakly dependent on the efficiency of atomic diffusion during the MS, as long as no extra
Li-burning is caused by the  
process invoked to mitigate the diffusion itself \citep[][herafter MSB12]{mucciarelli12}.\\
Also, the surface Fe abundance after the FDU is restored to essentially its initial value, because the iron
diffused from the convective envelope during the MS is almost completely re-engulfed by the deeper
convective layers. Hence, values of [Fe/H] measured in LRGB stars are representative of their initial Fe content.

In this work, we present measurements of \ali\ 
in a sample of LRGB stars, and perform two tests involving the progeny of dwarf stars on the Spite plateau and
in the {\sl meltdown} metallicity regime.
The aims are first to establish whether the turbulent diffusive chemical transport
proposed by \citet{ric02}, once calibrated to reproduce the Spite plateau starting from the SBBN A(Li), reproduces also \ali\ measured in LRGB stars, without any additional tuning; secondly, 
to assess whether LRGB stars display some signature of the {\sl meltdown} observed among dwarfs. This will help establishing whether this phenomenon is due to a real variation of the
initial \ali , or to some chemical transport process affecting the more metal-poor dwarfs.

\section{The LRGB sample}  
\label{ref}
We have collected a sample of high-resolution spectra of LRGB stars, combining proprietary 
(Program ID: 099.D-0287, PI: Mucciarelli,
CN2017A-33 and CN2017B-54, PI: Monaco) and archival data. 
Target stars were selected starting from the SAGA database \citep{suda08} and previous works focused
on metal-poor giant stars \citep[MSB12,][]{spite05,yong13,roederer14}.

\begin{figure}
\centering
\includegraphics[clip=true,width=0.5\textwidth]{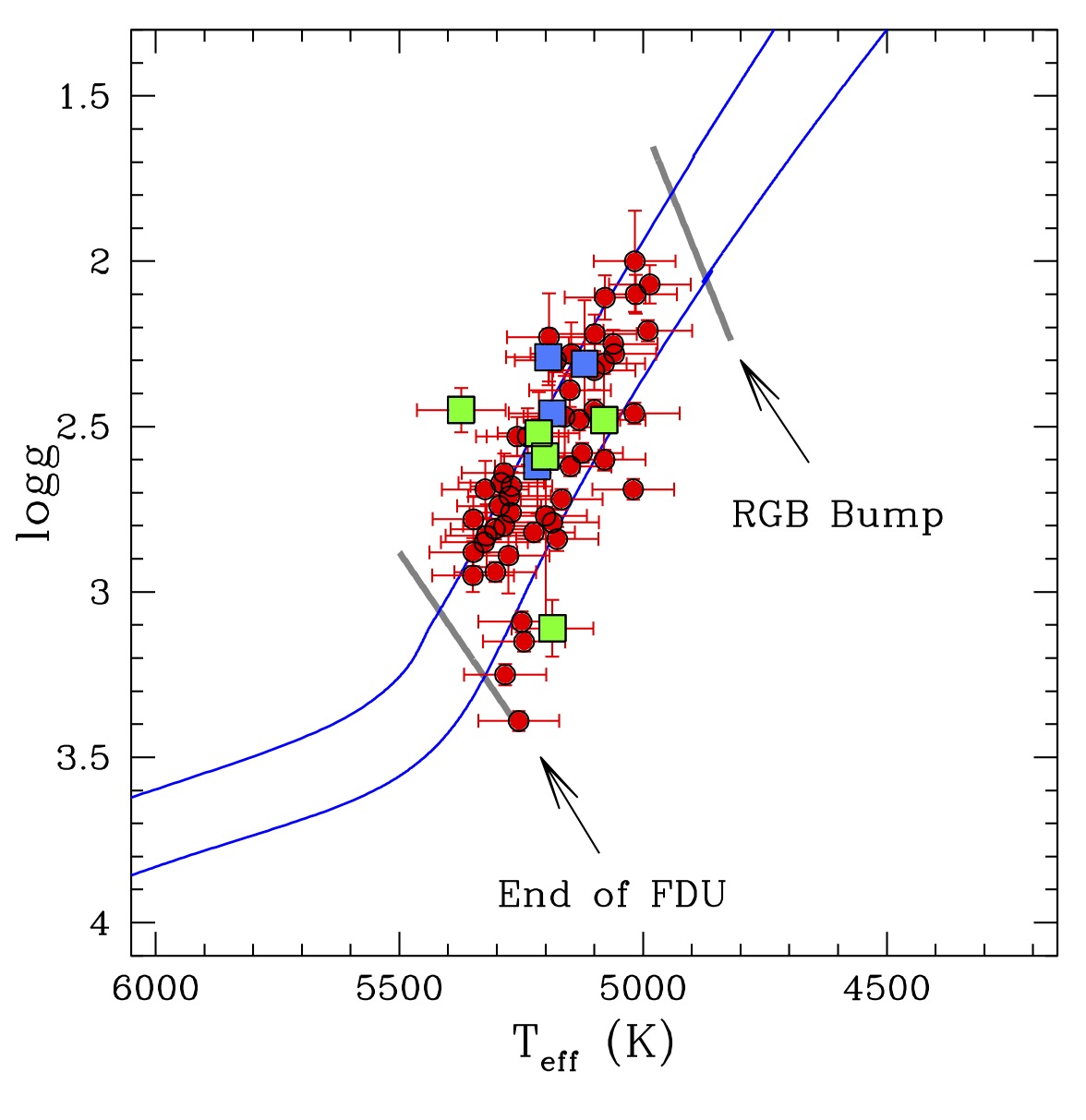}
\caption{Position of the observed LRGB targets (red circles) in the \teff\-logg diagram. 
Blue and green squares are low- and high-C band CEMP stars, respectively. 
Two BaSTI theoretical 
isochrones \citep{pietr21} with an age of 12.5 Gyr, 
[Fe/H]=--3.2 and --1.5 dex, and $\alpha$-enhanced chemical mixture are shown as reference (blue curves). 
The predicted positions of the FDU completion and 
the RGB-bump are marked as grey thick lines.}
\label{tg}
\end{figure}

The final choice of the targets is based on our homogeneous set of \teff\ and \gr\ (see Section~\ref{chems}), 
by selecting as {\sl bona-fide} LRGB stars those confined in a region of the \teff\--\gr\ diagram 
populated by RGB  models evolving between the end of the FDU and the RGB-bump (see MSB12), 
that satisfy the following criterion:
${\rm -0.00216 \cdot {\rm T_{\rm eff}} + 14.76 < \log{g} < -0.00366\cdot {\rm T_{\rm eff}} + 19.88}$ . 
The boundaries of this region are shown in Fig.~\ref{tg}. 
In addition, we have checked that these stars have not yet experienced the extra mixing 
episode occurring at the RGB-bump, by measuring their surface C and N abundances and, when possible, 
the C isotopic ratio \ciso\ 
(the latter can be measured in LRGB stars with [Fe/H]$> -$3.0 dex and in metal-poorer LRGB stars 
with high C abundance).  C abundances (available for all the stars) and N abundances (for two thirds of the targets), combined with \ciso\ higher than 10-15 (for half of the sample), confirm that the carbon-normal stars discussed here are LRGB stars.

The final sample includes 58 LRGB stars with [Fe/H]$< -$1.3 dex, 33 of them with [Fe/H]$< -$2.8 dex, corresponding to 
metallicities in the Li {\sl meltdown} region of dwarf stars (see Table 1); 
50 target stars have been observed with the spectrograph UVES at the Very Large Telescope of ESO \citep{dekker00}, 
4 stars with ELODIE mounted on the 1.93~m telescope of the Observatoire de Haute Provence \citep{Moultaka04}, 
3 stars with MIKE at the Magellan telescope \citep{bernstein03} and 1 star with HARPS at the ESO La Silla 3.6m telescope \citep{mayor03}.

Almost all spectra have resolution larger than 40,000 
 (in particular two thirds of the spectra used to measure the Li line have spectral resolution between 40,000 and 50,000) and signal-to-noise ratio per pixel larger than 100 around the Li line at 6708 \AA\ . Information about all the used spectra (instrument, spectral range and resolution, ID program) 
is listed in Table 2.
Compared to MSB12 we have enlarged the sample of LRGB stars by about a 
factor of three, and in particular we have now 19 stars with [Fe/H]$\leq -$3 dex, compared to only one star in MSB12.

%%%%%%%%%%%%%%%%%%%%%%%

\section{Chemical analysis}
\label{chems}
% stellar parameters
Stellar parameters have been obtained from the photometry of the early third data release of the ESA/$Gaia$ mission 
\citep{prusti16,brown20} in order to avoid the bias affecting the spectroscopic parameters at low metallicities \citep{mucciarelli20}.
\teff\ have been derived adopting the ${\rm (BP-RP)_0}$-\teff\  transformation by \citet{mbm21}.
Colour excesses are estimated from integration within 3D extinction maps \citep{ivanova21} but for the star HE0401-0138, 
for which we adopted the value by \citet{schlafly11}. For very distant halo stars falling out of the maps, 
we used the upper limits based on dust emission from \citet{planck16}, in all cases very close to the value achieved 
at the map boundary. The extinction coefficients 
have been derived adopting the iterative procedure described in \citet{lombardo21}.
Surface gravities have been obtained adopting the photometric \teff , a stellar mass equal to 0.8 ${\rm M_{\odot}}$, 
the parallaxes from Gaia and bolometric corrections
calculated from a grid of synthetic spectra computed with the code {\tt SYNTHE} \citep{kurucz05}. 
Microturbulent velocities have been estimated spectroscopically by minimising any trend between the abundances 
from Fe~I lines and their reduced equivalent widths.
%%%%%%

For all the targets we measured Fe and Li abundances (see Table~2). The Fe abundances have been derived with the code {\tt GALA} \citep{gala} 
from the measured equivalent widths \citep[the latter measured with {\tt DAOSPEC},][]{stetson08}. The Li abundances have been derived from spectral synthesis, by fitting the observed profile of the Li resonance line at 6708 \AA\ with suitable synthetic spectra.
3D-NLTE corrections from \citet{wang21} have been applied.
For 11 stars upper limits for \ali\  have been estimated by comparing the observed spectra with suitable grids of synthetic spectra computed with different values of \ali\ . The minimum \ali\ allowing to identify the Li line with respect to the observed noise has been evaluated through visual inspection. The similar level of the derived upper limits only reflects the similar spectral quality of the spectra and parameters of the stars.

Additionally, for all targets we measured the C abundance from the G-band or from the CH feature at 3143 \AA , to establish 
which objects are carbon enhanced metal poor (CEMP) stars. According to the selection criteria by \citet{bonifacio18}, we consider CEMP stars those stars 
with [C/Fe]$>$1.0 dex for [Fe/H]$>$-4.0 dex and 
with A(C)$>$5.5 dex for [Fe/H]$<$-4.0 dex. 
Among the CEMP stars, we distinguish between low-C band and high-C band stars, according to the two groups of stars usually identified in the A(C)-[Fe/H] diagram \citep{spite13}.

The total uncertainty on the measured abundances has been estimated by summing in quadrature the error in the
measurement procedure and that 
arising from the stellar parameters, including also the covariance terms. 
 The uncertainty arising from the fitting procedure (related to the quality of the spectra) 
has been estimated by means of Monte Carlo simulations, by creating for each star a sample of 
500 synthetic spectra with the appropriate instrumental resolution and pixel-step and adding Poissonian noise to reproduce the observed SNR. The line-fitting procedure has been repeated for these samples of simulated spectra, adopting as 1$\sigma$ uncertainty the dispersion of the  derived \ali\  distribution.

Radial velocities (RV) have been measured from the position of several metallic lines with {\tt DAOSPEC} with typical uncertainties lower than 0.1 ${\rm km~s^{-1}}$ .
We compared the new measured RVs (see Table 2) with those from previous studies 
\citep{cohen13,norris13,roederer14,arentsen19} and from the Gaia Data Release 2 \citep{gaiadr2}.  
For four targets only one RV measurement is available and we cannot discuss their possible binary nature.
Besides six targets for which RV variations have been already discussed in the literature, we
identified other five stars displaying RV variations which we consider belonging to binary systems (see Table 1).

\section{Results}

\begin{figure}%[!h]
\centering
\includegraphics[width=0.5\textwidth]{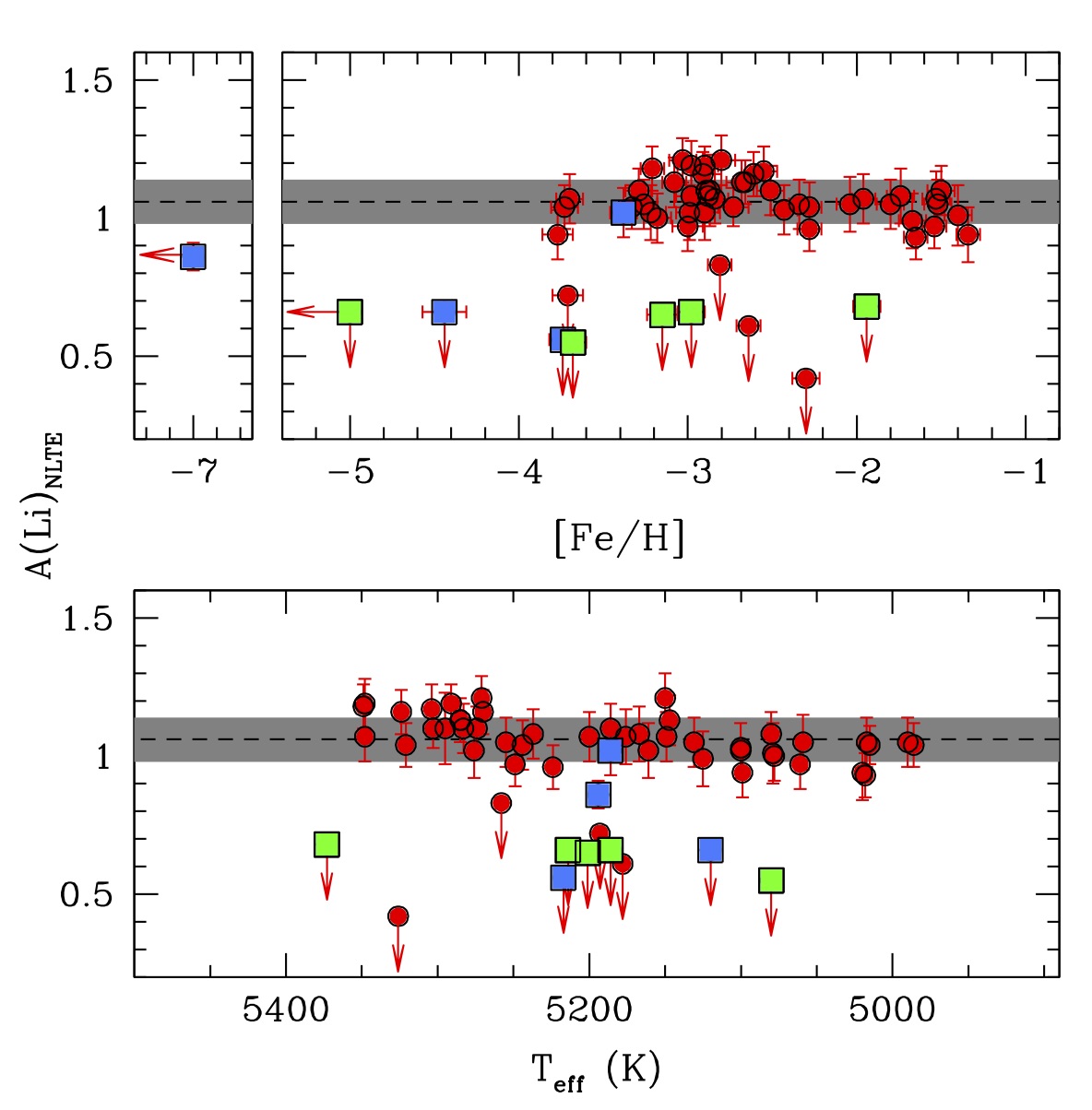}
\caption{Behaviour of A(Li) as a function of [Fe/H] and \teff\ 
(upper and lower panel, respectively). 
Symbols are the same of Fig.~\ref{tg}.
Arrows denote abundance upper limits. 
The dotted horizontal 
line denotes the average value of the measured A(Li), the grey shaded area indicates the $\pm 1\sigma$ region.}
\label{res}
\end{figure}

The behaviour of \ali\ as a function of [Fe/H] and \teff\ is shown in Fig.~\ref{res}. 
The distribution of \ali\ of the LRGB stars exhibits a clear dichotomy (thin plateau/Li-poor stars), with two well-separated groups of stars:\\ 
\begin{itemize}
\item {\sl Thin Li plateau}\\
For 47 stars the Li line is measurable, providing values of \ali\ very similar to each other, 
with an average  \ali\ =~1.06$\pm$0.01 dex ($\sigma$=~0.08 dex).
The star SMSS\_J031300.36-670839.3, the most metal-poor object of the sample ([Fe/H]$< -$7.0 dex), 
has  \ali\ =~0.86$\pm$0.05 dex, lower than the average value of the other stars. 
However, its extremely low iron content is outside the range of validity of the adopted colour-\teff\ relation.
If the star SMSS\_J031300.36-670839.3 is excluded, the average \ali\ of the sample 
is  1.07$\pm$0.01~dex ($\sigma$=~0.07 dex).
In both cases, the 1$\sigma$ dispersion is compatible with  an intrinsic null scatter, within
the uncertainties of \ali\ measurements, on the order of 0.08-0.09~dex, dominated by the uncertainty on \teff\ 
(on the order of 80-100 K). 

 We checked the statistical significance of the correlations between \ali\ and [Fe/H] and between \ali\ and \teff , calculating the Pearson linear correlation coefficient and the non-parametric Spearman and Kendall correlation coefficients. 
The derived p-values for A(Li) and [Fe/H] are 0.1, 0.35 and 0.32, above the standard 0.05 threshold.
This shows that the correlation between these two quantities is not statistically significant. On the other hand, the derived p-values for \ali\ and \teff are of about $10^{-5}$ for all the correlation coefficients. The measured slope between \ali\ and \teff\ is equal to 0.04$\pm$0.01 dex/100 K. 
Note that adoption of the 1D-NLTE corrections by \citet{lind09} erases this small slope, changing the average \ali\ only by 0.01 dex and not affecting our conclusions.

All the stars populating this Li plateau are carbon-normal stars, but
SMSS\_J031300.36-670839.3 and HE1506-0113 which are classified as low-carbon band CEMP stars. 
Eight stars display RV variations  (among them also the CEMP star HE1506-0113)
and they are considered as likely binary stars.

\item {\sl Li-poor stars}\\
For the remaining 11 stars the Li line is not detected and we can only provide upper limits for \ali , 
pointing out that these stars have \ali\ lower than 0.7 dex. 
This group of stars results to be highly heterogeneous, including seven CEMP stars and four carbon-normal stars. 
Among the CEMP stars, five of them are classified as high-C band and two as low-C band CEMP stars. 
We can evaluate possible RV variations for 10 Li-poor stars having multiple RV measurements. 
RV variations are identified for three (high-C band CEMP) stars of this group.
\end{itemize}

%%%%%%%%%%%%%%%%%%%%%%%

\section{Discussion}
\label{disc}

The distribution of \ali\ as a function of [Fe/H] for field halo LRGB stars is clearly different from that
of dwarfs, showing a clear dichotomy and no evidence of a {\sl meltdown}.

\begin{figure}
\centering
\includegraphics[width=0.5\textwidth]{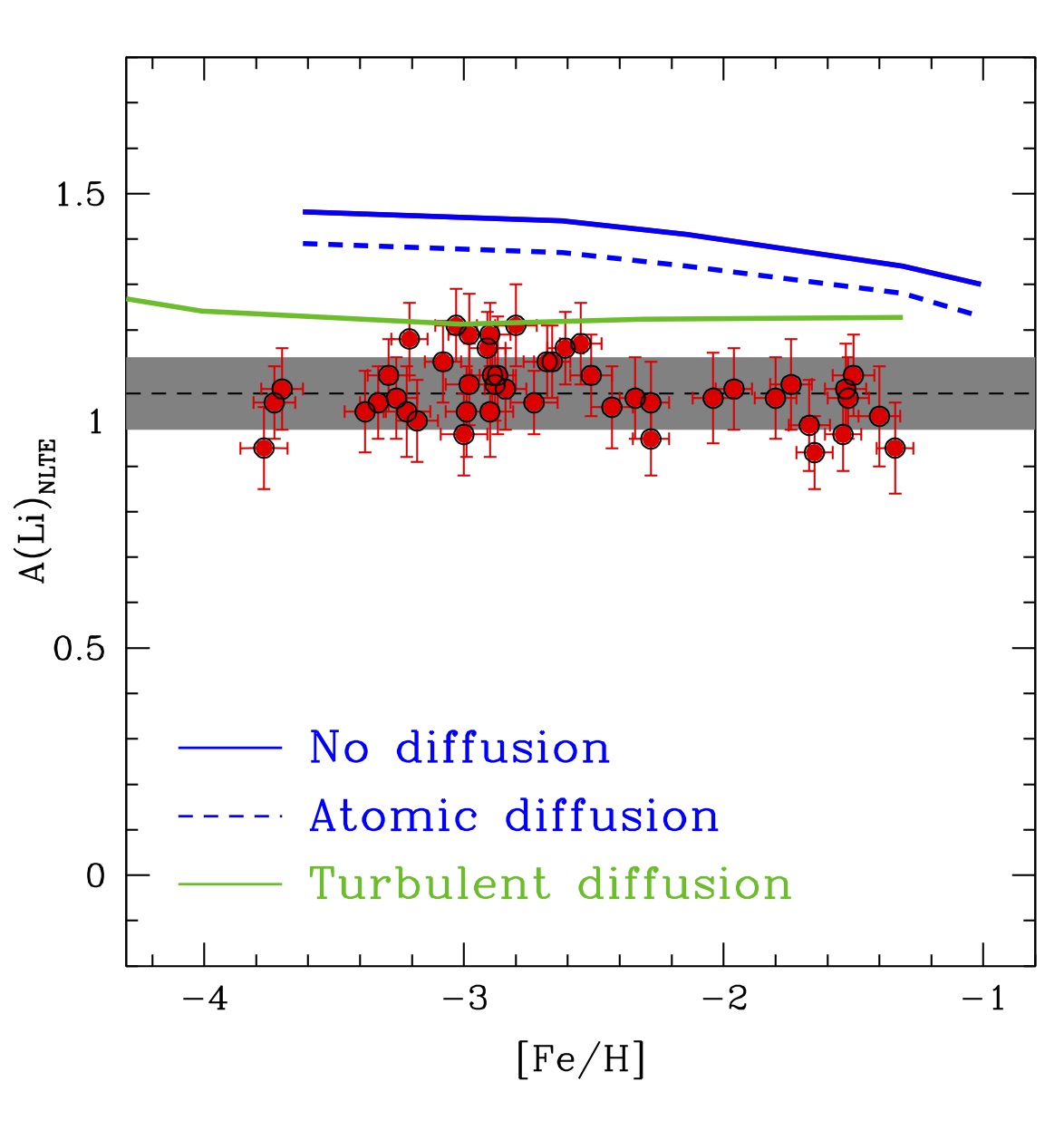}
\caption{Behaviour of \ali\ as a function of [Fe/H] for the stars on the RGB  Li plateau, compared
  to predictions from models calculated with and without atomic diffusion 
(blue dashed and solid curves, respectively) according to MSB12, and with atomic diffusion plus 
  an additional turbulent diffusion (green curve) as described in \citet{deal21}. All models are calculated with an initial SBBN \ali\ =~2.74 dex.}
\label{model}
\end{figure}

Figure~\ref{model} compares the measured values of \ali\ as a function of [Fe/H] for the LRGB stars on the Li plateau with predictions from standard theoretical models  (see MSB12 for details), assuming that all these stars formed with the SBBN abundance 
\ali\ =~2.74 dex \citep{coc17} regardless 
of their metallicity. We consider first models including only convection as mechanism of element transport 
(corresponding to a fully inhibited atomic diffusion), and models with also fully efficient atomic diffusion.

%%%%%%%%%%%%%%%%%%%%%%%

Both these sets of calculations are unable to reproduce the observed plateau, as  
they predict \ali\ values higher than the observed ones by 0.3-0.4~dex, and a mild decrease of A(Li)
with  increasing [Fe/H], because the FDU reaches deeper (in mass) layers when the metallicity increases.
An agreement with observations can be obtained only by assuming an initial A(Li) significantly lower (by 0.3-0.4 dex) than the SBBN value. Also, the initial A(Li)
should slowly increase with increasing [Fe/H], in stark contrast with 
current chemical evolution models that naturally predict 
a constant A(Li) for stars in this metallicity range \citep{matteucci21}.

To satisfy the constraint of a constant initial A(Li) equal to the SBBN value,  
some extra Li needs to be transported into the burning regions during the MS, 
to match the abundances measured in LRGB stars. 
The amount of this additional Li to be burned has to depend on the initial metallicity
of the stars, to counterbalance the variation of the post-FDU A(Li) with [Fe/H] predicted by the models. 

A parametrization of this additional transport process has been proposed by \citet{ric02, ric05} to explain the 
A(Li) measured in Spite plateau dwarfs in terms of the SBBN Li abundance. 
We consider here the stellar models calculated by \citet{deal21} including
\citet{ric02, ric05} parametrization,  and extended to the RGB phase.
In addition to convection and atomic diffusion, these models include a turbulent diffusion 
properly tuned to reproduce the Spite plateau.
These models (see Fig.~\ref{model}) predict a constant \ali\ for [Fe/H]$>$--4.0 dex, 
in agreement with our measurements, but with \ali\ about 0.15~dex higher than the measured values.
This relatively minor discrepancy may be accounted for by considering small uncertainties
in the theoretical predictions of \ali\ 
from SBBN calculations \citep[see e.g.][]{coc14} and systematics in the adopted \teff\ scale of LRGB stars. 

Some analyses of the BBN theory showed that the primordial lithium abundance could be smaller 
(by $\sim$0.1 dex) by considering the expected variation (few ppm) of fundamental constants 
\citep[see e.g. ][]{evans14,clara20,martins21,deal21b}. Such decrease of the primordial lithium abundance would 
improve the agreement between predicted and measured \ali\ .

It is remarkable that the calibration of this turbulent diffusion -- without any assumption on the physical 
causes of this mixing -- 
to reproduce theoretically the Spite plateau 
starting for the SBBN Li abundance, can also reproduce qualitatively the LRGB A(Li), despite a small offset.
The amount of extra Li burned on the MS by these models
is consistent with what we observe on the LRGB stars, despite the fact that the calibration of this 
turbulent diffusion is completely independent of LRGB observations.

Another important result of the comparison in Fig.~\ref{model} is the following.
Our LRGB sample defines a thin plateau over the entire range of [Fe/H] covered by the data, 
in particular at [Fe/H]$<$--2.8 dex, where a drop of the average A(Li) and
a large abundance dispersion is observed among the dwarf stars  \citep[see Figure 2 in][]{aguado19}. 
If these metal-poor dwarf stars really formed with a lower \ali\, we should expect the same proportion of Li-poor stars also among the LRGB stars. 
Only three out of 25 LRGB stars with [Fe/H]<–2.8 dex are Li-poor, 
whilst in the same metallicity range two-thirds of the dwarf stars with measured Li abundances have A(Li) below the Spite plateau. 
The lack of a significant dispersion of A(Li) on the LRGB down to an initial [Fe/H]$\sim -$3.8 dex seems to
rule out a spread of initial Li abundances to explain the Li {\sl meltdown} observed in dwarfs.

Given that the same parametrization of turbulent diffusion matching the Spite plateau
abundances in dwarfs can also explain the observed LRGB abundances down to an initial [Fe/H]$\sim -$3.8 dex,
the A(Li) values below the Spite plateau observed in {\sl meltdown} dwarf stars are likely due to yet another
surface chemical transport process efficient during the MS.
This process needs to be efficient, likely with star-to-star variations, only below a threshold
initial metallicity. Also, it must somehow deplete the surface \ali\ without
bringing extra Li to the burning region, otherwise we should observe its signature
in LRGB stars.

In addition to this mechanism, part of the A(Li) dispersion observed in the meltdown stars could arise from stars that have experienced thermohaline convection induced by binary mass transfer \citep{deal21}, or they can be the "blue-stragglers-to-be" proposed by  \citet{ryan01}, 
i.e. the result of the coalescence of two low-mass stars. The blue straggler nature of three out of four of
the Li-depleted stars studied by \citet{ryan01}
has been confirmed by their Gaia parallaxes \citep{bonifacio19}. 
The Li-poor LRGB stars that we observe are the progeny of these Li-poor stars populating 
the {\sl meltdown}.

Among the 11 Li-poor stars observed in our sample, five are high C-band CEMP stars and the lack 
of surface lithium can be easily explained as the result of mass transfer processes,
being most of these CEMP stars in binary systems \citep{hansen16,bonifacio18,arentsen19}. 
The other six Li-poor stars (two low-C band CEMP and four carbon normal stars) do not show evidence of RV variations and their low \ali\ cannot be explained by invoking mass transfer processes. These stars are 
good candidate to be blue straggler stars (i.e. the product of a coalescence process). However we cannot exclude other mechanisms able to form Li-poor stars.

When stars evolve along the RGB, the MS surface Li depletion due to the additional mixing process is erased by the deepening convection, which restores the uniform A(Li) as a function of [Fe/H] observed among the LRGB stars. On the other hand, the objects populating the meltdown that formed with A(Li) lower than the SBBN value, will show up as LRGB stars with a surface lithium abundance lower than the LRGB plateau.
This scenario qualitatively explains the existence of the {\sl meltdown} among the dwarfs and 
the \ali\ dichotomy among the giants.

To conclude, the discovery of a thin Li abundance plateau (coupled with a distinct Li-poor sub-population) in LRGB stars 
provides new insights to understand the evolution of A(Li) in metal-poor stars, and to constrain chemical
element transport mechanisms. 
The evidence coming from dwarf and LRGB metal poor stars suggest that they 
all formed with the same initial Li abundance consistent with the SBBN value, 
but those have experienced thermohaline convection induced by binary mass transfer 
or those formed from coalescence.
The turbulent diffusion proposed 
by \citet{ric02, ric05} seems to work well to explain dwarf and giant stars, reconciling 
the two sets of A(Li) measurements with the SBBN value.
As a note of caution, we stress that this is only a phenomenological solution to the {\sl cosmological lithium problem},
as we still lack a physical description of this process. 
Nevertheless, transport processes such as rotation-induced mixing and penetrative convection 
are promising candidates \citep{deal21b}.

Our analysis also shows that any future observational and theoretical investigation aimed at 
studying comprehensively the lithium abundances in metal-poor 
stars must be able to explain simultaneously the Spite plateau and {\sl meltdown} among dwarf stars, and 
the thin plateau among LRGB stars.

\tiny
\begin{table*}
\caption{Atmospheric parameters, iron, lithium, carbon abundances for the target stars. 
The SNR around the Li line is listed. The two last columns indicate whether the target star is classified as a CEMP 
(high=high-C band CEMP star, low=low-C band CEMP star, otherwise is a carbon-normal star) and as a binary (* = binary already identified, ** = binary from this work, ? = one only RV measurement, otherwise is a single star).
 }             % title of Table
\label{tab}      % is used to refer this table in the text
\centering                          % used for centering table
\begin{tabular}{l   c c c  c c c  c c c c}        % centered columns (4 columns)
\hline\hline                 % inserts double horizontal lines
ID &  \teff\ & \gr\ & $\xi$ & [Fe/H]  &  \ali\  &  [C/Fe]  &  SNR  & CEMP & Bin \\    % table heading 
\hline
   &  (K)   &  (cgs)  & (${\rm km~s^{-1}}$) & (dex) & (dex) &  (dex)   & (@6708) &  & \\
\hline                        % inserts single horizontal 

 BD+233130		        & 5304 &  2.81 &  1.3 &    -2.55$\pm$0.08 &  1.17$\pm$0.09   &    0.26$\pm$0.10   & 310   &    & \\
 BD-012582		        & 5224 &  2.82 &  1.2 &    -2.28$\pm$0.07 &  0.97$\pm$0.08   &    0.80$\pm$0.10   & 400   &    & \\
 BS 16467-062	                & 5348 &  2.78 &  1.6 &    -3.70$\pm$0.08 &  1.06$\pm$0.09   &    0.39$\pm$0.11   & 180   &    & \\
 BS 16477-003	                & 5017 &  2.00 &  1.9 &    -3.26$\pm$0.08 &  1.12$\pm$0.09   &    0.51$\pm$0.09   & 150   &    & ? \\
 CD-241782		        & 5285 &  2.80 &  1.0 &    -2.66$\pm$0.07 &  1.12$\pm$0.08   &    0.24$\pm$0.10   & 280   &    & \\
 CD-30298		        & 5274 &  2.71 &  0.6 &    -3.29$\pm$0.08 &  1.07$\pm$0.08   &    0.41$\pm$0.08   & 700   &    & \\
 CS 22183-031	                & 5271 &  2.76 &  1.0 &    -3.03$\pm$0.08 &  1.18$\pm$0.08   &    0.45$\pm$0.10   & 160   &    & ? \\
 CS 22186-023	                & 5178 &  2.30 &  1.4 &    -2.64$\pm$0.07 &  $<$0.65	     &    0.33$\pm$0.10   & 200   &    & \\
 CS 22877-001	                & 5061 &  2.25 &  1.2 &    -3.00$\pm$0.09 &  0.99$\pm$0.09   &    0.93$\pm$0.10   & 240   &    &  \\
 CS 22880-086	                & 5324 &  2.69 &  1.5 &    -2.91$\pm$0.07 &  1.16$\pm$0.08   &    0.72$\pm$0.13   & 120   &    & \\
 CS 22885-096	                & 5099 &  2.22 &  2.1 &    -3.77$\pm$0.09 &  0.98$\pm$0.09   &    0.46$\pm$0.10   & 230   &    &     \\
 CS 22896-154	                & 5285 &  2.64 &  1.5 &    -2.68$\pm$0.09 &  1.14$\pm$0.08   &    0.43$\pm$0.12   & 320   &    & ? \\
 CS 22897-008	                & 4986 &  2.07 &  2.1 &    -3.33$\pm$0.07 &  1.10$\pm$0.08   &    0.55$\pm$0.10   & 200   &    &     \\
 CS 22953-003	   	     	& 5150 &  2.39 &  1.6 &    -2.80$\pm$0.08 &  1.23$\pm$0.09   &    0.38$\pm$0.11   & 220   &    &     \\
 CS 22957-022	   	     	& 5258 &  2.53 &  1.1 &    -2.81$\pm$0.07 &  $<$0.84	     &    0.19$\pm$0.10   & 200   &    &    \\
 CS 29491-069	   	     	& 5270 &  2.68 &  1.2 &    -2.61$\pm$0.07 &  1.17$\pm$0.08   &    0.15$\pm$0.10   & 130   &    &     \\
 CS 29495-042	   	     	& 5373 &  2.45 &  1.0 &    -1.94$\pm$0.08 &  $<$0.73	     &    1.07$\pm$0.12   & 140   & high& **  \\
 CS 29502-042	   	     	& 5348 &  2.88 &  1.7 &    -2.98$\pm$0.08 &  1.16$\pm$0.09   &    0.37$\pm$0.10   & 360   &    &     \\
 CS 30312-100	   	     	& 5186 &  2.79 &  1.1 &    -2.51$\pm$0.08 &  1.10$\pm$0.09   &    0.29$\pm$0.12   & 140   &    & **  \\
 CS 31082-001	   	     	& 5015 &  2.10 &  1.8 &    -2.73$\pm$0.09 &  1.10$\pm$0.07   &    0.25$\pm$0.10   & 350   &    &     \\
 HD 002665		        & 5059 &  2.28 &  1.6 &    -2.04$\pm$0.08 &  1.14$\pm$0.10   &   -0.07$\pm$0.11   & 310   &    & \\
 HD 004306 		        & 5080 &  2.31 &  1.0 &    -2.88$\pm$0.08 &  1.10$\pm$0.08   &    0.46$\pm$0.10   & 450   &   & \\
 HD 006755 		        & 5176 &  2.84 &  1.4 &    -1.53$\pm$0.08 &  1.17$\pm$0.10   &   -0.03$\pm$0.09   & 160   &   & *   \\
 HD 021581		        & 5018 &  2.46 &  1.4 &    -1.65$\pm$0.07 &  1.04$\pm$0.08   &    0.11$\pm$0.08   & 140   &   & \\
 HD 026169		        & 5100 &  2.45 &  1.3 &    -2.43$\pm$0.08 &  1.07$\pm$0.09   &    0.32$\pm$0.10   & 700   &   & *   \\
 HD 027928		        & 5131 &  2.48 &  1.3 &    -2.34$\pm$0.06 &  1.09$\pm$0.09   &    0.12$\pm$0.12   & 450   &   & \\
 HD 045282		        & 5283 &  3.25 &  1.1 &    -1.50$\pm$0.08 &  1.17$\pm$0.09   &   -0.02$\pm$0.11   & 250   &   & \\
 HD 087140		        & 5167 &  2.72 &  1.4 &    -1.74$\pm$0.08 &  1.15$\pm$0.10   &    0.17$\pm$0.10   & 160   &   & \\
 HD 108317		        & 5326 &  2.85 &  1.2 &    -2.30$\pm$0.08 &  $<$0.42	     &    0.08$\pm$0.10   & 500   &	  &  \\
 HD 111721		        & 5020 &  2.69 &  1.2 &    -1.34$\pm$0.07 &  1.07$\pm$0.10   &   -0.05$\pm$0.11   & 400   &   & \\
 HD 126238		        & 4990 &  2.21 &  1.7 &    -1.80$\pm$0.08 &  1.16$\pm$0.09   &   -0.09$\pm$0.12   & 150   &    & \\
 HD 128279		        & 5244 &  3.15 &  1.0 &    -2.28$\pm$0.07 &  1.03$\pm$0.09   &   -0.04$\pm$0.11   & 330   &   & \\
 HD 175305		        & 5079 &  2.60 &  1.5 &    -1.40$\pm$0.08 &  1.13$\pm$0.11   &   -0.06$\pm$0.10   & 340   &    & \\
 HD 200654		        & 5303 &  2.94 &  0.8 &    -2.89$\pm$0.08 &  1.06$\pm$0.07   &    0.54$\pm$0.11   & 550   &    &  \\
 HD 211998		        & 5255 &  3.39 &  0.9 &    -1.52$\pm$0.08 &  1.11$\pm$0.09   &    0.12$\pm$0.09   & 540   &   & \\
 HD 218857		        & 5149 &  2.62 &  1.4 &    -1.96$\pm$0.07 &  1.13$\pm$0.09   &    0.04$\pm$0.10   & 320   &    & \\
 HD 220127		        & 5249 &  3.09 &  1.3 &    -1.54$\pm$0.07 &  1.04$\pm$0.08   &   -0.08$\pm$0.10   & 170   &    & \\
 HD 274939		        & 5125 &  2.58 &  1.4 &    -1.67$\pm$0.08 &  1.08$\pm$0.10   &    0.08$\pm$0.08   & 330   &    & \\
 HE 0037-2657		        & 5147 &  2.28 &  0.9 &    -3.08$\pm$0.07 &  1.18$\pm$0.09   &    0.36$\pm$0.10   & 180   &    & **  \\
 HE 0044-2459		        & 5349 &  2.95 &  0.8 &    -3.21$\pm$0.07 &  1.15$\pm$0.08   &    0.35$\pm$0.10   & 140   &    &    \\
 HE 0107-5240		        & 5214 &  2.52 &  1.5 &     $<$-5.00      &  $<$0.66	     &    4.48$\pm$0.13   & 130   & high& *	    \\
 HE 0132-2439		        & 5321 &  2.83 &  1.7 &    -3.73$\pm$0.08 &  1.00$\pm$0.08   &    0.72$\pm$0.10   & 200   &    & *  \\
 HE 0207-1423		        & 5186 &  3.11 &  1.3 &    -2.98$\pm$0.08 &  $<$0.63	     &    2.18$\pm$0.15   & 170   & high&  \\
 HE 0243-5238		        & 5295 &  2.74 &  1.1 &    -2.87$\pm$0.07 &  1.09$\pm$0.13   &    0.45$\pm$0.10   & 130   &    &     \\
 HE 0323-4529		        & 5237 &  2.53 &  1.0 &    -2.98$\pm$0.09 &  1.08$\pm$0.09   &    0.49$\pm$0.11   & 230   &    &     \\
 HE 0401-0138		        & 5078 &  2.11 &  0.9 &    -3.18$\pm$0.08 &  1.05$\pm$0.09   &    0.30$\pm$0.10   & 270   &    & **   \\
 HE 0547-4539		        & 5291 &  2.67 &  1.0 &    -2.90$\pm$0.08 &  1.18$\pm$0.07   &    0.58$\pm$0.10   & 140   &    &     \\
 HE 0557-4840		        & 5120 &  2.31 &  1.3 &    -4.44$\pm$0.13 &  $<$0.60	     &    1.77$\pm$0.15   &  70   &   low &     \\
 HE 0926-0546		        & 5193 &  2.23 &  1.7 &    -3.71$\pm$0.09 &  $<$0.70	     &    0.31$\pm$0.10   & 190   &   &     \\
 HE 1005-1439		        & 5201 &  2.59 &  1.5 &    -3.15$\pm$0.09 &  $<$0.64	     &    2.45$\pm$0.14   & 230   & high& ?     \\
 HE 1150-0428		        & 5080 &  2.48 &  1.4 &    -3.68$\pm$0.08 &  $<$0.55	     &    2.76$\pm$0.16   & 140   & high& *   \\
 HE 1219-0312		        & 5200 &  2.77 &  1.1 &    -2.84$\pm$0.08 &  1.06$\pm$0.09   &    0.03$\pm$0.10   & 170   &   &     \\
 HE 1347-1025		        & 5217 &  2.62 &  1.8 &    -3.74$\pm$0.08 &  $<$0.55	     &    1.06$\pm$0.10   & 190   &   low &     \\
 HE 1506-0113		        & 5186 &  2.46 &  1.5 &    -3.38$\pm$0.08 &  1.01$\pm$0.09   &    1.46$\pm$0.13   & 150   &   low & *   \\
 HE 2228-3806		        & 5276 &  2.89 &  1.0 &    -2.90$\pm$0.09 &  1.00$\pm$0.10   &    0.28$\pm$0.10   & 150   &    &     \\
 HE 2314-1554		        & 5161 &  2.47 &  1.1 &    -3.22$\pm$0.09 &  1.02$\pm$0.10   &    0.60$\pm$0.10   & 130   &    &     \\
 HE 2327-5642		        & 5100 &  2.33 &  1.1 &    -2.99$\pm$0.08 &  1.05$\pm$0.10   &    0.47$\pm$0.10   & 250   &    & **   \\
 SMSS J031300.36-670839.3       & 5194 &  2.29 &  1.5 &     $<$-7.00	  &  0.87$\pm$0.05   &    4.48$\pm$0.15   & 350   &   low &     \\

\hline                                   %inserts single line
\end{tabular}
\end{table*}

%%%%

\tiny
\begin{table*}
\caption{Instrument, spectral range and resolution, ID program and PI for each spectrum used in this work. RV is listed only for the stars for which we analysed spectra not already discussed in previous papers 
(the quoted errors are computed as the dispersion divided by the root mean square of the measured). The entire table is available in the electronic form.}             % title of Table
\label{tab}      % is used to refer this table in the text
\centering                          % used for centering table
\begin{tabular}{l   c c c  c c}        % centered columns (4 columns)
\hline\hline                 % inserts double horizontal lines
ID &  Instrument & Spectral range & Resolution &  RV  & ID Program (PI)  \\    % table heading 
\hline
   &     &  (\AA\ )  &  &(${\rm km~s^{-1}}$) &  \\
\hline                        % inserts single horizontal 

 BD+233130		        &  UVES   &  3281-4562   &   49620  &     --283.85$\pm$0.01 &	165.N-0276 (Cayrel)	\\	   
         		        &  UVES   &  4654-6760   &   51690  &     --283.80$\pm$0.02 &	165.N-0276 (Cayrel)	\\	   
 BD-012582		        &  UVES   &  3027-3884   &   40970  &     ---		    &	 68.D-0546 (Asplund)	 \\  % +1.43 0.03	<----- Garcia-Perez
           		        &  UVES   &  4729-6835   &   56990  &     ---		    &	 68.D-0546 (Asplund)	 \\  % +1.30 0.01
 BS 16467-062	                &  UVES   &  3305-4607   &   40970  &     ---		    &	165.N-0276 (Cayrel)	\\    
                                &  UVES   &  4654-6760   &   42310  &     ---		    &	165.N-0276 (Cayrel)	\\	    
 BS 16477-003	                &  UVES   &  3305-4607   &   40970  &     ---		    &	165.N-0276 (Cayrel)	\\    
                                &  UVES   &  4654-6760   &   42310  &     ---		    &	165.N-0276 (Cayrel)	\\    
 CD-241782		        &  UVES   &  3027-3884   &   40970  &     ---		    &	 68.D-0546 (Asplund)	 \\   %101.49  0.02   <----- Garcia-Perez
           		        &  UVES   &  4143-6277   &  107200  &     ---		    &	 69.D-0065 (Asplund)	 \\   %101.79  0.04
           		        &  UVES   &  4729-6835   &   56990  &     ---		    &	 68.D-0546 (Asplund)	 \\   %101.30  0.01
 CD-30298		        &  UVES   &  3027-3884   &   40970  &     ---		    &	 68.D-0546 (Asplund)	 \\  %28.09  0.02  <----- Garcia-Perez
           		        &  UVES   &  4729-6835   &   56990  &     ---		    &	 68.D-0546 (Asplund)	 \\  %27.72  0.02
 CS 22183-031	                &  UVES   &  3282-4563   &   40970  &    +16.73$\pm$0.03    &	099.D-0287 (Mucciarelli)  \\   
                                &  UVES   &  5655-9464   &   42310  &    ---   		    &	099.D-0287 (Mucciarelli)  \\  
 
\hline                                   %inserts single line
\end{tabular}
\end{table*}

\begin{acknowledgements}
 The authors thank the anonymous referee for the useful suggestions and E. Wang for her helpful comments about NLTE corrections.
  AM is grateful to the Scientific Council of Observatoire de Paris that funded his extended visit at GEPI, 
  where part of this work was carried out. 
  LM acknowledges support from {\sl proyecto interno} of the Universidad Andres Bello.
  MS acknowledges support from the STFC Consolidated Grant ST/V00087X/1.
  This work was supported by FCT/MCTES through the research grants 
  UIDB/04434/2020, UIDP/04434/2020 and PTDC/FIS-AST/30389/2017, and by FEDER - Fundo Europeu de 
  Desenvolvimento Regional through COMPETE2020 - Programa Operacional Competitividade e 
  Internacionaliza\c c$\tilde{a}$o (grant: POCI-01-0145-FEDER-030389). MD is supported by national funds through FCT 
  in the form of a work contract. OAR and MD acknowledge financial support from the 
  "Programme National de Physique Stellaire" (PNPS) of the CNRS/INSU co-funded by the CEA and 
  the CNES, France.
  
\end{acknowledgements}

%\clearpage
%-------------------------------------------------------------------

\end{document}